\documentclass{article}
\usepackage{amsmath,amssymb,amsthm}
\usepackage{cite}

\DeclareMathOperator{\diag}{diag}
\DeclareMathOperator*{\DS}{\oplus} 

\newcommand{\texbf}[1]{\text{\bf #1}}
\newcommand{\GL}{\texbf{GL}}

\newcommand{\real}{\mathbb{R}}
\newcommand{\comp}{\mathbb{C}}
\newcommand{\quat}{\mathbb{H}}
\newcommand{\A}{{\bf A}}
\newcommand{\Sm}{{\bf S}}
\newcommand{\IDEM}{\text{IDEM}_n(C^{p,q})}

\newcommand{\edn}{\end{document}}

\begin{document}
\title{Idempotents of Clifford Algebras\footnote{Presented at the 12\textsuperscript{th} International Colloquium ``Quantum Groups and Integrable Systems'', Prague, 12-14 June 2003.}} 
\author{R.\ Ab{\l}amowicz\thanks{rablamowicz@tntech.edu} \\
Department of Mathematics, Tennessee Technological University
\and
B.\ Fauser\thanks{Bertfried.Fauser@uni-konstanz.de}\\
Fachbereich Physik, Universit\"at Konstanz\and
K.\ Podlaski\thanks{podlaski@merlin.fic.uni.lodz.pl}, J.\ Rembieli\'{n}ski\thanks{jaremb@uni.lodz.pl}\\
Department of Theoretical Physics, University of {\L}\'{o}d\'{z}}
\maketitle

\noindent PACS: 02.10.Hh\\
{\em Key words}: Clifford algebras, idempotents, Jordan form, variety\\
{\tt Journal ref.}: Czech. J. Phys. {\bf 53}  (2003) 949

\begin{abstract}
A classification of idempotents in Clifford algebras $C^{p,q}$ is presented. It is shown that using isomorphisms between Clifford algebras $C^{p,q}$ and appropriate matrix rings, it is possible to classify idempotents in any Clifford algebra into continuous families. These families include primitive idempotents used to generate minimal one sided ideals in Clifford  algebras. Some low dimensional examples are discussed.
\end{abstract}

\section{Introduction}
In higher dimensional quantum field theories, as well as in higher dimensional relativistic quantum mechanics, a problem frequently arises how to project onto a specific invariant irreducible subspace of the spinor space. This problem is related to constructing and classifying idempotents of the corresponding Clifford algebras. Discrete families of primitive idempotents have been described by Chevalley \cite{[7]} and are traditionally used to generate spinor representations of Clifford algebras \cite{[8],[11]}.

In this paper we classify idempotents for arbitrary Clifford algebras $C^{p,q}$ over the reals. Some low dimensional examples are discussed.

In order to find all idempotent elements in a Clifford algebra $C^{p,q},$ we use its matrix representation. Due to the Wedderburn theorem, simple Clifford algebras are isomorphic to full matrix algebras over a division ring. The semisimple Clifford algebras are direct sums of such two simple algebras. In the Table~\ref{[table]} we recall the well known isomorphisms between Clifford algebras $C^{p,q}$ and matrix rings,
\begin{table}\caption{Isomorphisms of $C^{p,q}$ with matrix rings \cite{[1]}.}
\label{[table]}
\begin{center}
\begin{tabular}{ll}
  $p-q$ $(\text{mod } 8)=3,7$& $C^{p,q}\approx M(2^{\frac{p+q-1}{2}},\comp)$\\
  $p-q$ $(\text{mod } 8)=2,0$& $C^{p,q}\approx M(2^{\frac{p+q}{2}},\real)$\\
  $p-q$ $(\text{mod } 8)=4,6$& $C^{p,q}\approx M(2^{\frac{p+q-2}{2}},\quat)$\\
  $p-q$ $(\text{mod } 8)=1$& $C^{p,q}\approx M(2^{\frac{p+q-1}{2}},\real)\DS M(2^{\frac{p+q-1}{2}},\real)$\\
  $p-q$ $(\text{mod } 8)=5$& $C^{p,q}\approx M(2^{\frac{p+q-3}{2}},\quat)\oplus M(2^{\frac{p+q-3}{2}},\quat)$
\end{tabular}
\end{center}
\end{table}
\noindent
where $\real,\comp,\quat$ denote the real numbers, the complex numbers and the quaternions, respectively. 

In the following we find all classes of idempotents, up to an isomorphy of the chosen basis, using a specific matrix representation for every particular Clifford algebra $C^{p,q}$.    

\section{The case $p-q$ $(\text{mod } 8)=3,7$}\label{s1}

Let us start with case $C^{p,q}\approx M(2^{\frac{p+q-1}{2}},\comp)$. Suppose that  $\A\in M(n,\comp)$ with $N=2^\frac{p+q-1}{2}$ is an idempotent matrix. By means of  a similarity transformation it can be transformed to its Jordan form. Consequently, the idempotency implies that the Jordan form of $\A$ must be (up to a transposition of the basis vectors) of the form $\Pi_n=\diag\{\underbrace{1,1,\ldots,1}_n,\underbrace{0,\ldots,0}_{N-n}\}$,
$n=0,1,2,\ldots,N$; note two trivial cases: $n=0$ ($\A=0$) and $n=N$ ($\A=I$).  Therefore, $\A$ has the form
\begin{equation*}
\A=\Sm\Pi_n\Sm^{-1}\quad \text{where} \quad \Sm\in{\bf G}=\GL(N,\comp)\,.
\end{equation*} 
Evidently, the families --i.e. orbits of the $\GL(N,\comp)$ group action-- of idempotents can be distinguished by the trace of their elements, since these traces are group characters. Each orbit of an idempotent family is generated 
from a single $\Pi_n$. The $\Pi_n$ form a section of the discrete Weyl group, generating the ring of irreducible representations. The stability group of $\Pi_n$ is the direct product \mbox{${\bf G}_o:=\GL(n,\comp)\times\GL(N-n,\comp)$}. 
The matrices $\Sm$ can be decomposed as $\Sm_{{\bf G}/{{\bf G}_o}}\Sm_{{\bf G}_o}$ with $\Sm_{{\bf G}_o}\in {\bf G}_o$ and $\Sm_{{\bf G}/{{\bf G}_o}}\in {\bf G}/{\bf G}_o$, $[{\bf G}_o\,\Sm]=0$. It is clear that
\begin{equation*}
\A=\Sm_{{\bf G}/{\bf G}_o}\Pi_n\Sm_{{\bf G}/{\bf G}_o}^{-1},
\end{equation*}
i.e., the manifold of the idempotents in this case forms a variety homeomorphic to the coset space
\begin{equation*}
{\bf G}/{\bf G}_o=\frac{\GL(N,\comp)}{\GL(n,\comp)\times\GL(N-n,\comp)}.
\end{equation*}
Denoting this manifold by $\IDEM,$ we obtain the following families of idempotents for $(p-q) \, \text{mod }8=3,7:$
\begin{equation*}
\IDEM\approx \frac{\GL(N,\comp)}{ \GL(n,\comp)\times\GL(N-n,\comp)}.
\end{equation*}
where $N=2^\frac{p+q-1}{2}$, $n=0,1,2,\ldots,N$.

Notice that the real dimension of $\text{IDEM}_n$ in these cases is
\begin{equation*}
\dim_\real \IDEM=2(N^2-n^2-(N-n)^2)=4n(N-n)\,.
\end{equation*}

\section{The case $p-q$ $(\text{mod } 8)=2,0$}\label{s2}

In this case $C^{p,q}\approx M(2^\frac{p+q}{2},\real)$. In complete analogy to the previous case we obtain the following classification of idempotents:
\begin{equation*}
\IDEM\approx \frac{\GL(N,\real)}{\GL(n,\real)\times\GL(N-n,\real)},
\end{equation*}
where $N=2^\frac{p+q}{2}$, $n=0,1,2,\ldots,N$. In particular, 
\begin{equation*}
\dim_\real \IDEM=2n(N-n).
\end{equation*}

\section{The case $p-q$ $(\text{mod } 8)=4,6$}\label{s3}

In this case $C^{p,q}\approx M(2^\frac{p+q}{2},\quat)$. As before\footnote{The problem of diagonalization of quaternionic matrices and their Jordan form was considered, for example, in \cite{[2],[3],[4]}.} we easily conclude that in this case
\begin{equation*}
\IDEM\approx \frac{\GL(N,\quat)}{\GL(n,\quat)\times\GL(N-n,\quat)}.
\end{equation*}
$N=2^\frac{p+q-2}{2}$, $n=0,2,4,\ldots,N$, and\footnote{$n$ is even because of the preservation of the quaternionic structure.}
\begin{equation*}
\dim_\real \IDEM=8n(N-n).
\end{equation*}

\section{The case  $p-q$ $(\text{mod } 8)=1$}\label{s4}

In this case $C^{p,q}=M(2^\frac{2+q-1}{2},\real)\DS M(2^\frac{2+q-1}{2},\real)$. Taking into account idempotency condition
we can diagonalize idempotent $\A$ in this case to the form $\Pi_{n,m}=\diag\{\underbrace{1,1,\ldots,1}_n,\underbrace{0,\ldots,0}_{N-n};
\underbrace{1,1,\ldots,1}_m,\underbrace{0,\ldots,0}_{N-m}\}$ where $N=2^\frac{p+q-1}{2}$. The stability group of $\Pi_{n,m}$ should be a subgroup of $\GL(N,\real)\times\GL(N,\real)$ so it is simply
$(\GL(n,\real)\times\GL(N-n,\real))\times(\GL(m,\real)\times\GL(N-m,\real))$.
Therefore, the families of idempotents in this case are homeomorphic to the coset space
\begin{equation*}
\text{IDEM}_{n,m}(C^{p,q})\approx \frac{\GL(N,\real)}{\GL(n,\real)\times
\GL(N-n,\real)}\times\frac{\GL(N,\real)}{\GL(m,\real)\times\GL(N-m,\real)}\,,
\end{equation*}
$N=2^\frac{p+q-1}{2}$, $n=0,1,\ldots, N$, $m=0,1,\ldots, N$, and
\begin{equation*}
\dim_\real \text{IDEM}_{n,m}(C^{p,q})=2(n(N-n)+m(N-m))\,.
\end{equation*}

\section{The case  $p-q$ $(\text{mod } 8)=5$}\label{s5}
In this case $C^{p,q}=M(2^\frac{2+q-3}{2},\quat)\DS
M(2^\frac{2+q-3}{2},\quat)$. Using results obtained in Sect.\ref{s3} we find  that
\begin{equation*}
\text{IDEM}_{n,m}(C^{p,q})\approx \frac{\GL(N,\quat)}{\GL(n,\quat)\times\GL(N-n,\quat)}\times\frac{\GL(N,\quat)}{\GL(m,\quat)\times\GL(N-m,\quat)}\,,
\end{equation*}
$N=2^\frac{p+q-3}{2}$, $n=0,2,4,\ldots, N$, $m=0,2,4,\ldots, N$, and
\begin{equation*}
\dim_\real \text{IDEM}_{n,m}(C^{p,q})=8(n(N-n)+m(N-m))\,.
\end{equation*}

\section{Low dimensional examples}\label{ex}

\subsection{The case $p-q$ $(\text{mod } 8)=3,7$}\label{ex1}

The lowest dimensional examples in this case are $C^{3,0}$ and $C^{1,2}$. We have chosen the following basis for these algebras:
\begin{equation*}
\left.
\begin{array}{l}
C^{3,0}=\text{span}\{I,\vec{\sigma},i\vec{\sigma},iI\}\\[0.5ex]
C^{1,2}=\text{span}\{I,\sigma^1,i\sigma^2,i\sigma^3,-i\sigma^1,\sigma^2,-\sigma^3,iI\}\\
\end{array}
\right\}\sim M(2,\comp),
\end{equation*}
where the $\sigma_i$'s are Pauli matrices. Thus
\begin{equation*}
\A=\frac{1}{2}(I+\vec{\varphi}\vec{\sigma}),\qquad \vec{\varphi}=\vec{y}+i\vec{z}\quad        
\text{ where }\vec{y},\vec{z}\in\real^3 \,.
\end{equation*}
If $\A$ is an idempotent matrix, then we obtain two equations for $\vec{y},\vec{z}$ 
of the form:
\begin{gather}
\vec{y}^2-\vec{z}^2=1, \label{hyper33}\\
\vec{y}\vec{z}=0\,.\label{cone}
\end{gather}

We see that the hypersurface given by \eqref{hyper33} can be identified with the hyperboloid $H_{3,3}$ in $\real^6$. It can be shown that \eqref{cone} represents a rotated cone $V_{3,3}$. Therefore, the variety of idempotents in the Clifford algebras $C^{3,0}$ and $C^{1,2}$ are isomorphic to the intersection of the hyperboloid \eqref{hyper33} and the cone \eqref{cone},
$H_{3,3}\cap V_{3,3}$.
\begin{gather*}
\text{IDEM}_1(C^{3,0})=\text{IDEM}_1(C^{1,2})=\frac{\GL(2,\comp)}{
\GL(1,\comp)\times\GL(1,\comp)}\approx H_{3,3}\cap V_{3,3}\,,\\[0.5ex]
\dim_\real \text{IDEM}_1(C^{3,0})=\dim\text{IDEM}_1(C^{1,2})=4*1(2-1)=4\,.
\end{gather*} 
Now we can display exact expressions for the most general idempotent element $\A$ in our examples.
\begin{itemize}
\item In $C^{3,0}$, if we denote basis elements as
\begin{gather*}
e^0=I, \quad e^1=\sigma^1, \quad e^2=\sigma^2, \quad e^3=\sigma^3,\\
e^{12}=i\sigma^1, \quad e^{23}=i\sigma^2,\quad e^{31}=i\sigma^3,\quad e^{123}=iI\,,
\end{gather*}
then 
\begin{gather*}
\A=\frac{1}{2}(e^0+y_1e^1+y_2e^2+y_3e^3+z_1e^{12}+z_2e^{23}+z_3e^{31})\,.
\end{gather*}
\item In $C^{1,2}$, if we denote basis elements as
\begin{gather*}
e^0=I,\quad e^1=\sigma^1,\quad e^2=i\sigma^2,\quad e^3=i\sigma^3,\\
e^{12}=-i\sigma^1,\quad e^{23}=\sigma^2,\quad e^{31}=-\sigma^3,\quad e^{123}=iI\,,
\end{gather*}
then we find 
\begin{gather*}
\A=\frac{1}{2}(e^0+y_1e^1+z_2e^2+z_3e^3-z_1e^{12}+y_2e^{23}-y_3e^{31})\,.
\end{gather*}
In both cases, $\vec{y}$ and $\vec{z}$ satisfy \eqref{hyper33} and \eqref{cone}.
\end{itemize}

\subsection{The case $p-q$ $(\text{mod } 8)=2,0$}\label{ex2}

The lowest dimensional Clifford algebras in this case are $C^{2,0}$ and $C^{1,1}$. We express all elements of these algebras in the following basis:
\begin{gather}
\left.
\begin{array}{l}
C^{2,0}=\text{span}\{I,\sigma^3,\sigma^1,i\sigma^2\}\\[0.5ex]
C^{1,1}=\text{span}\{I,\sigma^1,i\sigma^2,-\sigma^3\}\\
\end{array}
\right\}\sim M(2,\real)\,, \nonumber \\[0.5ex]
\A=\frac{1}{2}(I+\vec{\varphi}\vec{\sigma}),\qquad \vec{\varphi}=(x,iz,y) \quad \text{ where }x,y,z\in\real\,, 
\nonumber \\[0.5ex]
\A^2=\A\Rightarrow\vec{\varphi}^2=1\,, \nonumber\\[0.5ex]
\vec{\varphi}^2=1\Rightarrow x^2+y^2-z^2=1\,.\label{hyper21}
\end{gather}

If $\A$ is an idempotent matrix, then we find that $\{\vec{\varphi}^2\}$ defines a one-sheet hyperboloid $H_{2,1}$ 
given by Eq. \eqref{hyper21}. Therefore the variety of idempotents $\text{IDEM}_1(C^{2,0})$ and $\text{IDEM}_1(C^{1,1})$ is isomorphic to such a hyperboloid.
\[
\text{IDEM}_1(C^{2,0})=\text{IDEM}_1(C^{1,1})=\frac{\GL(2,\real)}{\GL(1,\real)\times\GL(1,\real)}\approx H_{2,1}
\]
Note also that $\dim_\real \text{IDEM}_1(C^{2,0})=\dim_\real \text{IDEM}_1(C^{1,1})
=2*1(2-1)=2$. We can rewrite the idempotent matrices in the chosen basis for the Clifford algebras under study.
\begin{itemize}
\item In $C^{2,0}$ we choose a basis  
\begin{gather*}
e^0=I,\quad e^1=\sigma^3,\quad e^2=\sigma^1,\quad e^{12}=i\sigma^2\,,
\end{gather*}
and obtain
\begin{gather*}
\A=\frac{1}{2}(e^0+ye^1+xe^2+ze^{12})\,.
\end{gather*}
\item In $C^{1,1}$ we choose a basis 
\begin{gather*}
e^0=I,\quad e^1=\sigma^1,\quad e^2=i\sigma^2,\quad e^{12}=-\sigma^3\,,
\end{gather*}
and  obtain
\begin{gather*}
\A=\frac{1}{2}(e^0+xe^1+ze^2-ye^{12})\,.
\end{gather*}
\end{itemize}
\noindent
In the both cases, $x,$ $y,$ and $z$ satisfy equation \eqref{hyper21}.

For concrete low dimensional calculations it is convenient to use computer algebra systems.
The presented low dimensional examples can be performed using {\tt CLIFFORD}\footnote{http://www.math.tntech.edu/rafal/}, a package for 
Maple\footnote{http://www.maplesoft.com} \cite{[5],[6]}.

\section{Conclusions}

We have demonstrated that idempotents of Clifford algebras form algebraic varieties. These are smooth manifolds in the natural topology. Since idempotents can be used to generate left and right spinor ideals, which carry representations of the algebra under consideration, these varieties contain valuable information about invariant subspaces and the representations. Since Clifford algebras can in general be obtained by tensoring lower dimensional Clifford algebras, present work provides information about tensor product representations that is valuable for branching processes and possibly for quantum information processing.

An obvious task for future research is to study intertwiners between the idempotent orbits. This is related to topological questions about the Witt ring on representations and should then be related to Brauer-Wall groups. One might expect
to gain insight via, for example, Hopf fibration of spheres into the number and mutual relations of the intertwiners. Furthermore, it may be promising to reinspect the Radon-Hurwitz number involved in the matrix algebra isomorphisms, which counts the number of global vector fields, by performing the present work directly in an algebraic setting without using the matrix representations.


\begin{thebibliography}{9}
\bibitem{[7]} C.~Chevalley: {\em The Algebraic Theory of Spinors\/}, Columbia University Press, New York, 1954.
\bibitem{[8]} P.~Lounesto, {\em Clifford Algebras and Spinors\/}, 2nd ed., Cambridge University Press, Cambridge, May 2001.
\bibitem{[11]} R.~Ab{\l}amowicz: Comp. Phys. Comm. Thematic Issue Computer Algebra in Physics Research, {\bf 115} (1998) 510.
\bibitem{[1]} J.~Rembieli\'{n}ski: in {\it Clifford Algebras and their Applications in Mathematical 
Physics} (Eds. A.~Micali \textit{et al.,}), Kluwer Academic Publishers, 1992, p. 97.
\bibitem{[2]} H.~Aslaksen: Math. Intel., {\bf 18} 57 (1996).
\bibitem{[3]} De~S.~Leo and G.~Scolarici: J. Phys. A {\bf 33} (2000) 2971.		 
\bibitem{[4]} N.~Cohen  and De~S.~Leo: Electronic J. Linear Algebra. {\bf 7} (2000) 100.
\bibitem{[5]} R.~Ab{\l}amowicz, P.~Lounesto and J.M.~Parra: {\it Clifford Algebras with Numeric and Symbolic 
Computations}, Birkh\"{a}user, Boston, 1996.
\bibitem{[6]} R.~Ab{\l}amowicz and B.~Fauser: {\it Clifford Algebras and their Applications 
in Mathematical Physics} (Vol 1), Birkh\"{a}user, Boston 2000.
\end{thebibliography}
\end{document}